\documentclass[9pt,twocolumn,twoside]{osajnl}
\usepackage{mathtools}
\usepackage{amsmath}
\usepackage{widetext}
\usepackage{accents}
\journal{pr} 

\setboolean{shortarticle}{false}

\title{Disorder-Protected Quantum State Transmission through Helical Coupled-Resonator Waveguides}

\author[1,2]{JungYun Han}
\author[3]{Andrey A. Sukhorukov}
\author[1,2,*]{Daniel Leykam}

\affil[1]{Center for Theoretical Physics of Complex Systems, Institute for Basic Science, Daejeon 34126, Korea.}
\affil[2]{Basic Science Program, University of Science and Technology, Daejeon 34113, Korea.}
\affil[3]{Nonlinear Physics Centre, Research School of Physics, The Australian National University, Canberra, ACT 2601, Australia}

\affil[*]{Corresponding author: dleykam@ibs.re.kr}

\begin{abstract}
We predict the preservation of temporal indistinguishability of photons propagating through helical coupled-resonator optical waveguides (H-CROWs). H-CROWs exhibit a pseudospin-momentum locked dispersion, which we show suppresses onsite disorder-induced backscattering and group velocity fluctuations. We simulate numerically the propagation of two-photon wavepackets, demonstrating that they exhibit almost perfect Hong-Ou-Mandel dip visibility and then can preserve their quantum coherence even in the presence of moderate disorder, in contrast to regular CROWs which are highly sensitive to disorder. As indistinguishability is the most fundamental resource of quantum information processing, H-CROWs may find applications for the implementation of robust optical links and delay lines in the emerging quantum photonic communication and computational platforms.

\end{abstract}

\setboolean{displaycopyright}{true}

\begin{document}

\maketitle
\section{Introduction}
Topological photonics is emerging as a way to create disorder-immune waveguides for light using edge modes of media with non-trivial topological properties~\cite{Haldane08, Lu14,Ozawa19}. Since the first proof-of-concept experiments using gyro-magnetic microwave photonic crystals~\cite{Wang08,Wang09}, various approaches have emerged to demonstrate topological transport at different length and energy scales~\cite{Lu14,Ozawa19}. Moreover, this field is developing by harmonizing with existing sub-fields of photonics, i.e. exploring the role of topology in nonlinear optical effects~\cite{Smirnova19}, dynamically-modulated systems~\cite{Ozawa19Syn}, lasers~\cite{Harari18,Miguel18,Ota20}, and other non-Hermitian systems with structured gain and loss~\cite{Gong18}.

\begin{figure}
\centering
\includegraphics[width=0.6\columnwidth,height=8.0cm, angle =-90]{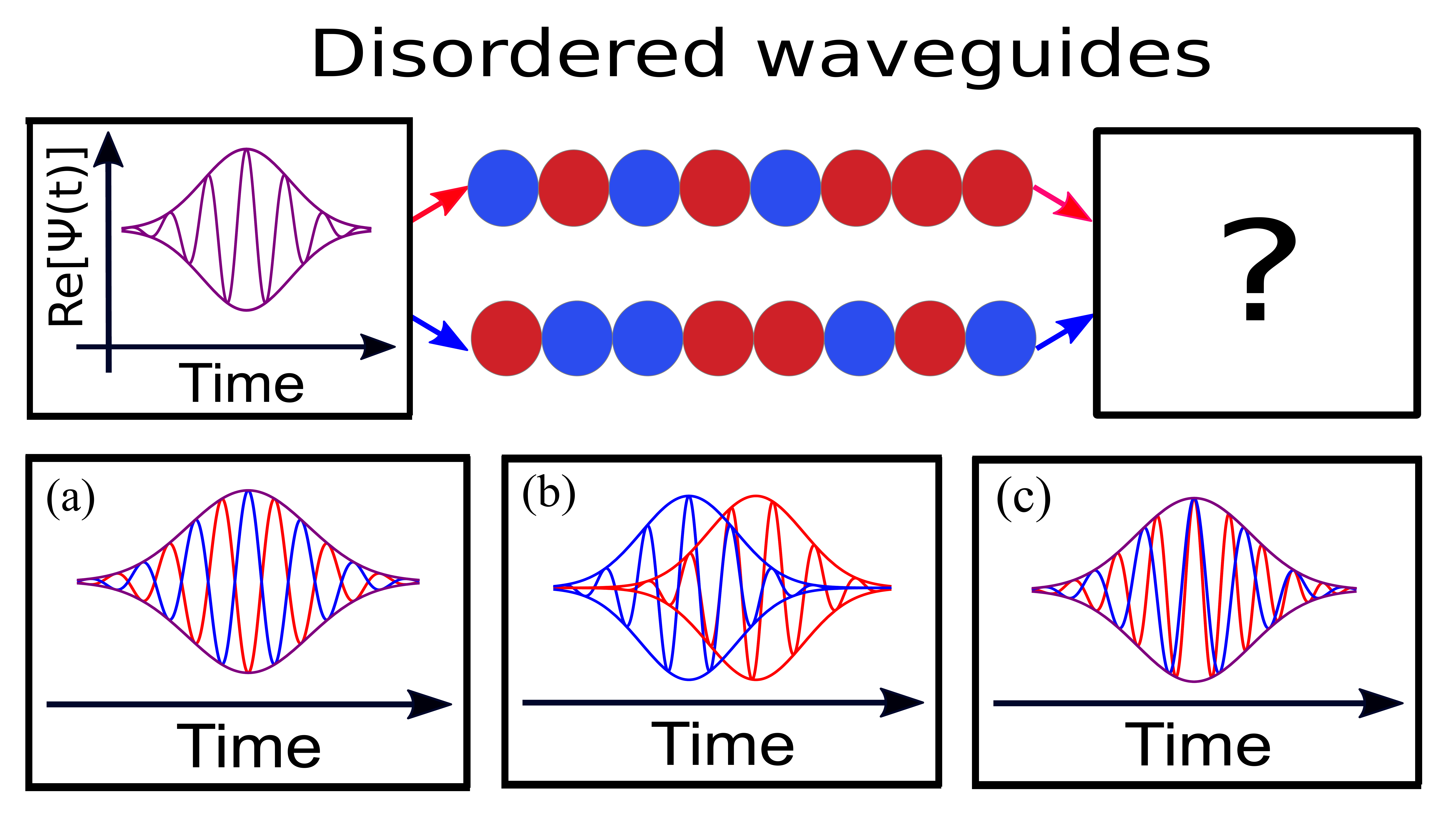}
\caption{Temporally indistinguishable photons within the temporal resolution $\delta t$ propagating through different delay lines can be temporally distinguishable given the delay provided by the ring resonator waveguides is sensitive to disorder, i.e. random red or blue shifts of the individual resonators. Insets below illustrate various possible effects of disorder: (a) Phase shift via the difference in phase velocities, (b) difference in arrival times due to variation of the group velocities, and (c) wavepacket distortion due to higher-order dispersion and wavelength-dependent reflection.}
\label{fig:indistinguishability}
\end{figure}

One promising application of topological photonics is the robust generation and transport of quantum state of light~\cite{Mittal18,Tambasco18,Blanco-Redondo18,Wang19}. It is a more challenging problem, as the preservation of quantum properties such as indistinguishability depends on the phase information, which is often \emph{not} robust against disorder even for topological edge states~\cite{Gneiting17}. Protecting indistinguishability has turned out to be very important as it provides a desirable resource in quantum technologies to generate entanglement~\cite{Streltsov17}. In optical delay lines, indistinguishability is determined by the degree of temporal overlap, as depicted in Fig.~\ref{fig:indistinguishability}. An obstacle to preservation of temporal overlap through standard delay lines is disorder, which can not only induce backscattering and Anderson localization of propagating waves but also destroys information related to the relative phase, which occurs due to the dephasing of the ensemble-averaged state~\cite{joos13,Kropf16}.
 
A potential solution is to use the robustness of topological edge states to protect quantum states of light~\cite{Rechtsman16,Mittal16E,Tambasco18,Clemens19}. For example, photonic quantum spin-Hall phases host bi-directional edge states which are protected against backscattering by time-reversal and internal (e.g. crystalline) symmetries. Quantum spin-Hall edge states were demonstrated in two-dimensional silicon ring resonator arrays, where the circulation direction within the rings plays the role of spin~\cite{Hafezi11,Hafezi13,Mittal14,Mittal16}. Other approaches introduce sublattices as a pseudospin degree of freedom emulating real spins~\cite{Wu15,Plotnik16}. The quantum spin-Hall phase supports transport robust against certain types of disorder, as backscattering requires a spin flip. 
 
So far, topologically-protected waveguiding has been largely demonstrated in two or higher dimensional photonic systems, while 1D topologically-protected transport has required synthetic dimensions~\cite{Lustig19} or time modulation such as adiabatic pumping~\cite{Kraus12}. There is another approach to achieve disorder-resistant waveguiding in 1D without a topological bandgap, based on directly implementing a spin-momentum locked dispersion by breaking time-reversal symmetry, which has been demonstrated using 1D electronic quantum wires~\cite{Quay10}. 
 Recently, we proposed a model of a quasi-1D coupled-resonator optical waveguide (CROW) exhibiting a similar helical spin-momentum locked dispersion (H-CROW), by combining circulation direction and sublattice spin-like degrees of freedom. The former effectively breaks time-reversal symmetry, while the coupling between different sublattices can be tuned to create a sublattice-momentum locked dispersion relation~\cite{Han19}.
This results in a suppression of backscattering and enhancement of localization length compared to the regular CROW model, as well as preservation of phase information. It is noteworthy as it provides the way to miniaturize disorder-resistant waveguide by inducing helical transport.
 
In this manuscript we study the propagation of quantum states of light through H-CROWs, and demonstrate that temporal indistinguishability can be robust against moderate disorder, thereby enabling the protection of entangled states. First, in Sec.~\ref{sec:model} we compute the delay time distribution of single-photon states propagating through disordered H-CROWs, and find that for a given disorder strength, H-CROWs yield a narrower distribution of delays compared to regular CROWs. Thus, the temporal overlap of photons travelling along different paths can be preserved. In Sec.~\ref{sec:indisting} we compute coincidence probability as a witness of indistinguishability and show the Hong-Ou-Mandel (HOM) dip~\cite{Hong88}. We then study in Sec.~\ref{sec:entang} the propagation of a path-entangled photonic state (N00N state), and show that H-CROWs protect their entanglement. We quantify their purity, associated with an inverse HOM dip, and obtain larger entanglement entropy compared to regular CROWs. We present conclusions and outlook in Sec.~\ref{sec:concl}.

\section{Model of H-CROWs with disorder-robust transport} \label{sec:model}
\begin{figure}
\centering
\includegraphics[width=\columnwidth]{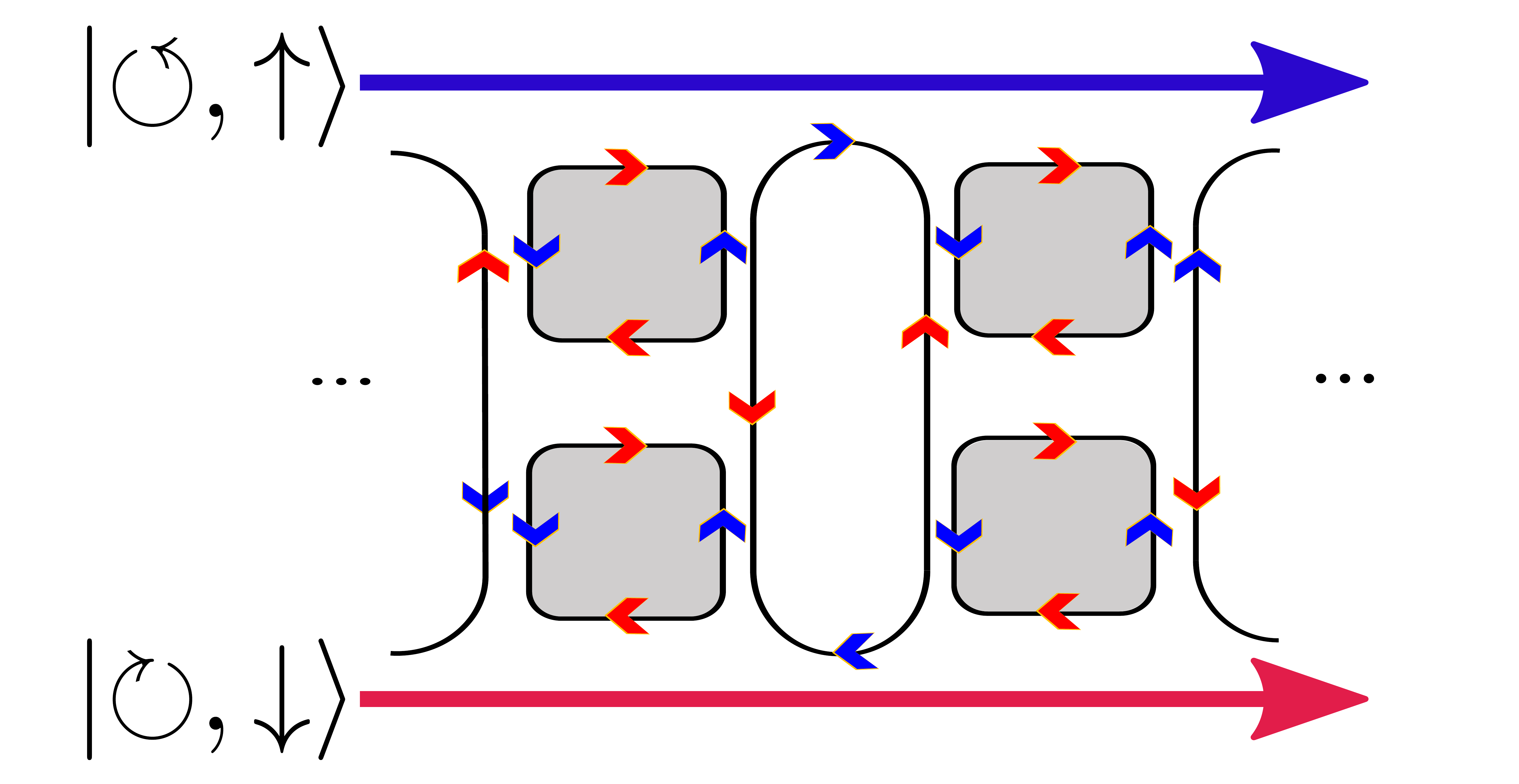}
\caption{Schematic of the helical coupled-resonator optical waveguide (H-CROW). Pseudospin-momentum locking is achieved after a certain propagation distance, where each sublattice exhibits definite momentum for designated circulation mode, thereby facilitating a disorder-resistant transport.
As opposite circulations exhibit opposite helicity, two co-propagating channels can be realized.}
\label{fig:model}
\end{figure}

The helical-CROW (H-CROW) consists of two legs of resonant ring cavities coupled via off-resonant link rings, as illustrated in Fig.~\ref{fig:model}. We assume decoupled circulation modes of each cavity, which allows one to effectively break time-reversal symmetry by making the link rings asymmetric. We are interested in the propagation of wavepackets close to the band center, where the tight-binding approximation is valid~\cite{yariv99,Hafezi11}. Under the tight-binding approximation, each unit cell hosts two sites, with coupling between nearest and next-nearest neighbors. 

  The tight-binding Hamiltonian governing time evolution of one specific circulation direction (counter-clockwise) in the absence of disorder reads~\cite{Hafezi11,Hafezi13,Leykam18,Han19},
  \begin{equation}
\begin{aligned}
&\hat{H}_{0,\text{ccw}} = \sum_{n} \left(\hat{H}_{a,\text{ccw}} + \hat{H}_{b,\text{ccw}} + \hat{H}_{ab,\text{ccw}} + \hat{H}^{\dagger}_{ab,\text{ccw}}\right), \\
&\hat{H}_{a,\text{ccw}} = J\hat{a}^{\dagger}_{n}\left( -i \hat{a}_{n-1} + i \hat{a}_{n+1} \right), \\
&\hat{H}_{b,\text{ccw}} = J\hat{b}^{\dagger}_{n}\left( i \hat{b}_{n-1}  -i \hat{b}_{n+1} \right),\\
&\hat{H}_{ab,\text{ccw}} = 2J \hat{a}^{\dagger}_{n}\left(\hat{b}_{n} + \frac{1}{2}\left( \hat{b}_{n-1} + \hat{b}_{n+1}\right)  \right),
\end{aligned}
	\label{eq:Hamiltonian}   	
\end{equation}
where $J$ is the hopping strength and field operators $\hat{a}_{n},\hat{b}_n$ represent the annihilation operators at the sites in the $n$th unit cell, while their conjugates are associated with the corresponding creation operators. The complex hopping terms arise due to the specially introduced asymmetry of the link rings~\cite{Han19}. Note that we measure frequencies with respect to a resonance of a single isolated ring, such that the eigenvalues of $\hat{H}_{0,\text{ccw}}$ are modal frequency detunings with respect to this resonance.

Let us now consider a periodic lattice, which enables us to compactly write the Hamiltonian in $k$-space as
\begin{equation}
\hat{H}_{0,\text{ccw}} =\sum_k \psi_{k,\text{ccw}}^{\dagger} \left(  \boldsymbol{d}(k) \cdot \boldsymbol{\hat{\sigma}} \right) \psi_{k,\text{ccw}},
    \label{eq:Hamiltonian matrix}
\end{equation}
where $k$ is the crystal momentum, $\psi_{k,\text{ccw}} = (\hat{a}_{k,\text{ccw}},\hat{b}_{k,\text{ccw}})^T$, $\textbf{d} = (d_{0},d_{x},d_{y},d_{z}) = (0,2J(1+\cos k),0,-2J\sin k)$, $\boldsymbol{\hat{\sigma}}$ are Pauli matrices, and 
we now interpret the upper ($\hat{a}$) and lower ($\hat{b}$) layers as corresponding to up and down pseudospin degrees of freedom respectively. 
The eigenvalues of $\hat{H}_{0,\text{ccw}}$ are $\omega_{\pm}(k) =  \pm2\sqrt{2}J  \sqrt{1+\cos{k}}$. The first component, $d_{0}$, describes the symmetric part of the intra-leg coupling which vanishes under our choice of hopping phase~\cite{Han19}, $2J\hat{\sigma}_x $ is analogous to a Zeeman field, and $ 2J\cos k \hat{\sigma}_x $ and $-2J \sin k \hat{\sigma}_z$ resemble intrinsic and Rashba-like spin-orbit couplings, respectively.

For the opposite excitation (clockwise), propagation is governed by the time-reversed Hamiltonian $\hat{H}_{\text{cw}}$ which exhibits opposite hopping phases due to time reversal symmetry~\cite{Gottfried03}. We introduce a total Hamiltonian composed of both circulations given in the direct product form, $\hat{H}_{0,\text{tot}} = \hat{H}_{\text{ccw}} \oplus \hat{H}_{\text{cw}}$. We obtain pseudospin-momentum locking in the center of the pass-band ($k = \pi$), where $\omega = 0$, $d_{x} = 0$, and the wave group velocity becomes $d\omega_{\pm} / dk = \pm 2J$. It supports the most resistant light propagation against disorder, since $\mathcal{H}_{0,\text{tot}}(k)$ for the small momentum deviations $k = \pi + \Delta k$ has the form
\begin{equation}
    \mathcal{H}_{0}(\pi + \Delta k) \approx  2J \ \mathrm{diag}( 	\Delta k, -\Delta k , -\Delta k , \Delta k ),
\label{eq:Hamiltonian momentum}
\end{equation}
where~\eqref{eq:Hamiltonian momentum} is written in the basis $\lbrace\vert$$\circlearrowright,\uparrow \rangle, \vert$$ \circlearrowright,\downarrow \rangle, \vert $$\circlearrowleft,\downarrow \rangle , \vert $$\circlearrowleft,\uparrow \rangle \rbrace$; the first index labels the circulation direction and the second indicates the pseudospins. Note that the off-diagonal component describing pseudospin-flipping vanishes in the first (linear) order of deviation.

At the band center H-CROWs show the maximum Anderson localization length and most resistant temporal pulse propagation, since the most significant disorder is misalignment of the rings' resonant frequencies, which is diagonal in the sublattice basis and does not flip the pseudospin~\cite{Canciamilla10, Hafezi13, Mittal14, Mittal18, Han19,Gneiting17}.

Importantly, we can employ the circulation degree of freedom to use H-CROWs as two-mode delay lines, see Fig.~\ref{fig:model}. By exciting both circulations through different sublattices, the simultaneous pseudospin-momentum locking phases (red and blue) can be obtained. We now consider the propagation of light in the presence of disorder and losses. For the sake of simplicity, we only take the dominant onsite disorder into account, which has the form~\cite{Han19} 
\begin{equation}
\hat{V}_{\epsilon} = \sum_n \left( V^{(a)}_{n,\epsilon} \hat{a}_n^{\dagger} \hat{a}_n + V^{(b)}_{n,\epsilon} \hat{b}_n^{\dagger} \hat{b}_n \right), \label{eq:disorder}
\end{equation}    
where $\epsilon$ labels each disorder realization. We assume that each on-site potential $V^{(j)}_{n,\epsilon}$ has a Gaussian distribution with mean zero and standard deviation $U$.
 
We first formulate equations for the field operators $\hat{\psi}_{n} = (\hat{a}_n,\hat{b}_n)^T$ to calculate the transmission at a given frequency $\omega$, the equation reads~\cite{Gardiner85,Hafezi13}
\begin{equation}
\begin{aligned}
    i \omega\hat{\psi}_{n,j}(\omega) =& i \left[\hat{H}_{0,j},\hat{\psi}_{n,j}(\omega)\right] -\kappa_{\text{ex}}\hat{\psi}_{n,j}(\omega) (\delta_{n,1}+\delta_{n,L})  \\&
    -\kappa_{\text{in}} \hat{\psi}_{n,j}(\omega)
    + \sqrt{2 \kappa_{\text{ex}}}\ \hat{p}_{\text{in},j}(\omega)\delta_{n,1},
\label{eq:Langevin equation frequency}
\end{aligned}
\end{equation}
where $j = 1$ (ccw) and $j = 2$ (cw) index counter-clockwise and clockwise circulation modes, respectively, $\kappa_{\text{in}}$ are the intrinsic scattering losses of each cavity, $\kappa_{\text{ex}}$ is coupling strength to the input/output leads at each edge of the array, and $L$ is the array length. The input field operator $\hat{p}_{\text{in,j}}(\omega)$ is defined by
\begin{equation}
    \hat{p}_{\text{in,1}}(\omega) = p_{\text{in}}(\omega)\hat{a}_{n}(\omega)
    ,\quad \hat{p}_{\text{in,2}}(\omega) = p_{\text{in}}(\omega)\hat{b}_{n}(\omega),
\end{equation}
corresponding to wavepackets with \emph{identical} temporal distributions but opposite circulations and sublattices, which excite the two helical modes.

The reflection (\textit{R}) and transmission (\textit{T}) amplitudes can be expressed for the two inputs as follows~\cite{Hafezi11,Hafezi13},
\begin{equation}
\begin{aligned}
  &R_{1}(\omega) = \left\vert \frac{p_{\text{in}}(\omega)-\sqrt{2\kappa_{\text{ex}}}a_{1}(\omega)}{p_{\text{in}}(\omega)}  \right\vert^{2}, \ T_{1}(\omega) = \left\vert \frac{\sqrt{2\kappa_{\text{ex}}}a_{L}(\omega)}{p_{\text{in}}(\omega)}  \right\vert^{2}, \\&R_{2}(\omega) = \left\vert \frac{p_{\text{in}}(\omega)-\sqrt{2\kappa_{\text{ex}}}b_{1}(\omega)}{p_{\text{in}}(\omega)}  \right\vert^{2}, \ T_{2}(\omega) = \left\vert \frac{\sqrt{2\kappa_{\text{ex}}}b_{L}(\omega)}{p_{\text{in}}(\omega)}  \right\vert^{2},
\label{eq:reflection and transmission}
\end{aligned}
\end{equation}
where operators without a 'hat' refer to their corresponding field components of sublattices $a,b$~\cite{Schwabl97}. The derivation is given in Appendix~\ref{app:in-out}. We note the following relation for the  output field operators ~\cite{Gardiner85},
\begin{equation}
\hat{p}_{\text{out,1}} = -\sqrt{2\kappa_{\text{ex}}}\hat{a}_{L}, \quad \hat{p}_{\text{out,2}} = -\sqrt{2\kappa_{\text{ex}}}\hat{b}_{L}.
\end{equation}

\begin{figure}
\centering
\includegraphics[width=\columnwidth,height=9cm]{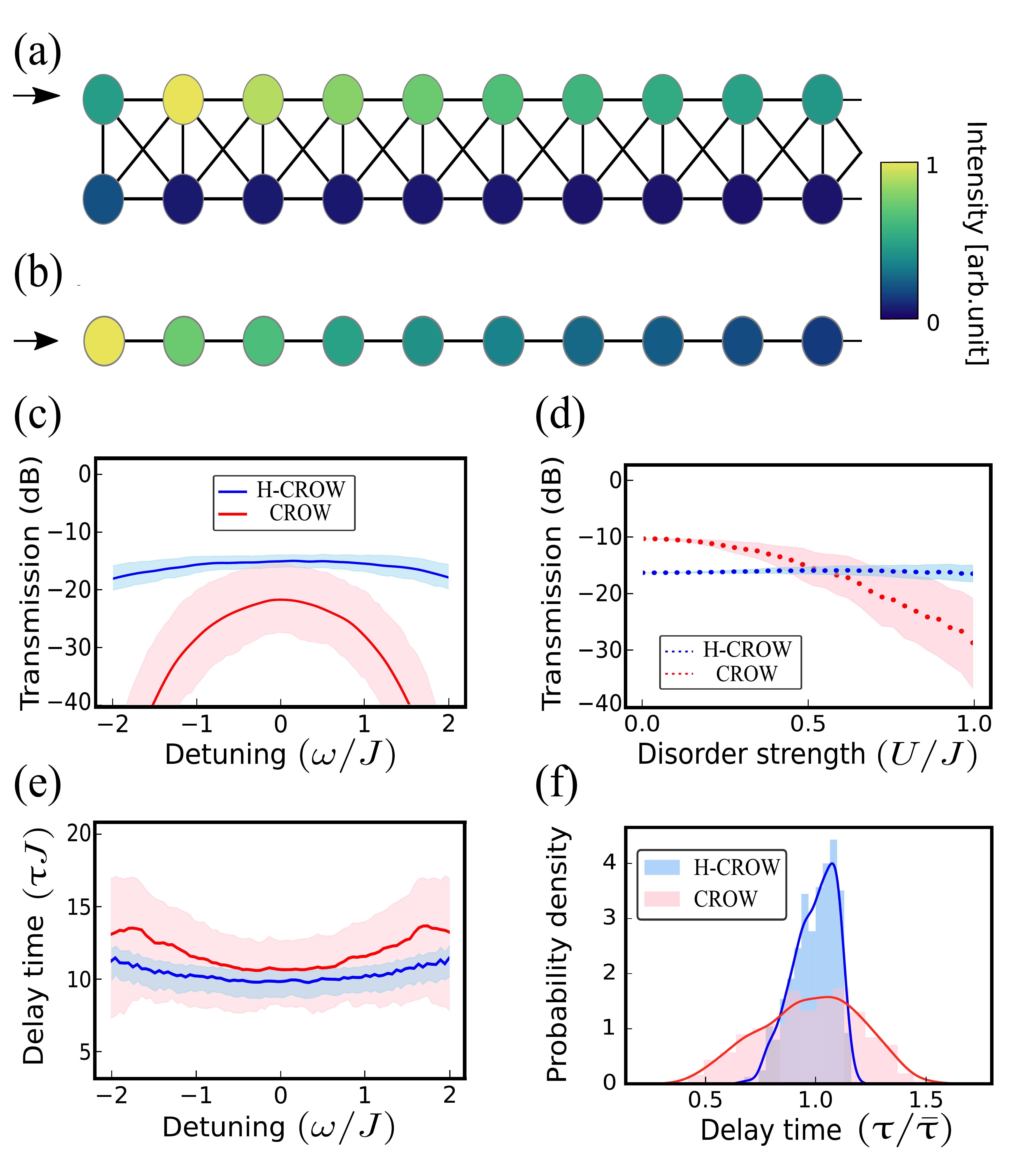}
\caption{Classical wave transport through H-CROWs and CROWs in the presence of moderate disorder $U=0.8J$ and intrinsic losses $k_{\text{in}} = 0.1J$. (a,b) Disorder-averaged field intensity profiles at $\omega = 0$ in the first ten rings of an $L=20$ H-CROW (a) and CROW (b). (c) Frequency-dependent transmission spectra. Solid lines indicate the disorder average and shaded regions represent $65\%$ confidence interval. Maximum of average is -15.8 dB and -22.8dB at $\omega = 0$ for the H-CROW and CROW, respectively. (d) Dependence of the transmission at $\omega = 0$ on the disorder strength $U$. (e) Wavepacket delay time as a function of the input frequency. (f) Distribution of delay times at $\omega = 0$, where $\bar{\tau}$ is the root mean square delay.}
\label{fig:profiles}
\end{figure}

Given the transmission spectrum, we can compute the wavepackets' group delay times $\tau_j$ via
\begin{equation}
    \tau_{j} = \frac{1}{i}\frac{d}{d\omega}\left(\frac{p_{\text{out},j}(\omega)}{\vert p_{\text{out},j}(\omega)\vert}\right) \quad (j=1,2),
\label{eq:group delay}
\end{equation}
where $p_{\text{out,j}}(\omega)$ refers to the corresponding field component. This quantity measures the transit time of a wave packet through the device. The distribution of delay times provides a measure of the sensitivity of the system to disorder~\cite{Mittal14}.

We present in Fig.~\ref{fig:profiles}(a) disorder-averaged intensity profiles in the first half of an $L=20$ H-CROW for a monochromatic input at $\omega = 0$ (the middle of the transmission band). For comparison, Fig.~\ref{fig:profiles}(b) shows the intensity profile of a regular CROW. The numerical calculations are performed using parameters similar to the experiments of Refs.~\cite{Hafezi13,Mittal14}: $\kappa_{\text{ex}} = 0.5J$, $k_{\text{in}} = 0.1J$, disorder standard deviation $U = 0.8J$, and an ensemble of $500$ realizations. In the H-CROW, the intensity remains confined to the upper ($a$) sublattice, signifying the pseudospin-momentum locking, with the attenuation of the intensity along the array occurring only due to the internal scattering losses $\kappa_{\mathrm{in}}$. 
 The regular CROW also exhibits Anderson localization, resulting in more rapid attenuation of the intensity in Fig.~\ref{fig:profiles}(b). 
 
We plot in Fig.~\ref{fig:profiles}(c) the transmission spectra~\eqref{eq:reflection and transmission}, where the shaded regions indicate the $65\%$ confidence intervals~\cite{Hafezi13}. H-CROWs achieve higher transmission than regular CROWs through the entire passband, with almost five times higher transmission at $\omega = 0$ and smaller deviations between different disorder realizations as a result of the disorder resistant uni-directional propagation. Fig.~\ref{fig:profiles}(d) demonstrates that H-CROWs maintain the amplitude of the output field for a broad range of disorder strengths $U$, while CROWs shows increasingly poor performance as $U$ increases. Based on these results, we use $U = 0.8J$ as a disorder strength for the following studies, as this regime clearly illustrates the advantage of H-CROWs vs. regular CROWs in presence of the moderate disorder typically present in experiments~\cite{Hafezi11}. We note that the comparatively lower transmission of the H-CROW for a weak disorder is due to a non-optimal input coupling, which can be improved by e.g. coupling the input waveguide to the second unit cell of the array. 

The effects of disorder on the group delay time vs. frequency detuning are presented in Fig.~\ref{fig:profiles}(e). We see that H-CROWs exhibit a slightly smaller average delay, but much lower fluctuations. Furthermore, the statistics of delay times in the vicinity of $\omega = 0$ presented in Fig.~\ref{fig:profiles}(e) show that H-CROWs exhibit more ballistic transport than regular CROWs, as for H-CROWs the delay time distribution around the root mean square average has Gaussian shape with smaller variance.

\section{Preservation of photon indistinguishability} \label{sec:indisting}

In this section, we consider the transmission of two identical photons forming a separable quantum state at the input, and analyze the degree of temporal photon indistinguishability at the output of the H-CROW and CROW.
Specifically, we consider an input state $\vert 11\rangle_{\text{cw,ccw}}$, with one photon in the clockwise mode and a temporally identical photon in the counter-clockwise mode. The output state in the frequency domain has the form

\begin{equation}
\begin{aligned}
    &\vert\text{out}\rangle = \hat{\phi}_{a}\hat{\phi}_{b}\vert 00\rangle_{ab} \\&= \int d\omega \int d\omega' p_{\text{out},1}(\omega) p_{\text{out},2}(\omega') \hat{a}^{\dagger}_{\text{out}}(\omega)\hat{b}^{\dagger}_{\text{out}}(\omega')\vert 00\rangle_{ab},    
\end{aligned}
\end{equation}
where the subscripts $a$ and $b$ indicate the Hilbert space corresponding to upper/lower part of output port with the field operators $\hat{\phi}_{a/b} \coloneqq \int d\omega \hat{p}_{\text{out},1/2}(\omega)$ composed of field creation operators of each output port $\hat{a}^{\dagger}_{out},\hat{b}^{\dagger}_{out}$. Note that we are working with scalar fields, assuming a fixed polarization state.

\begin{figure}
\centering
\includegraphics[width=\columnwidth,height=8.0cm]{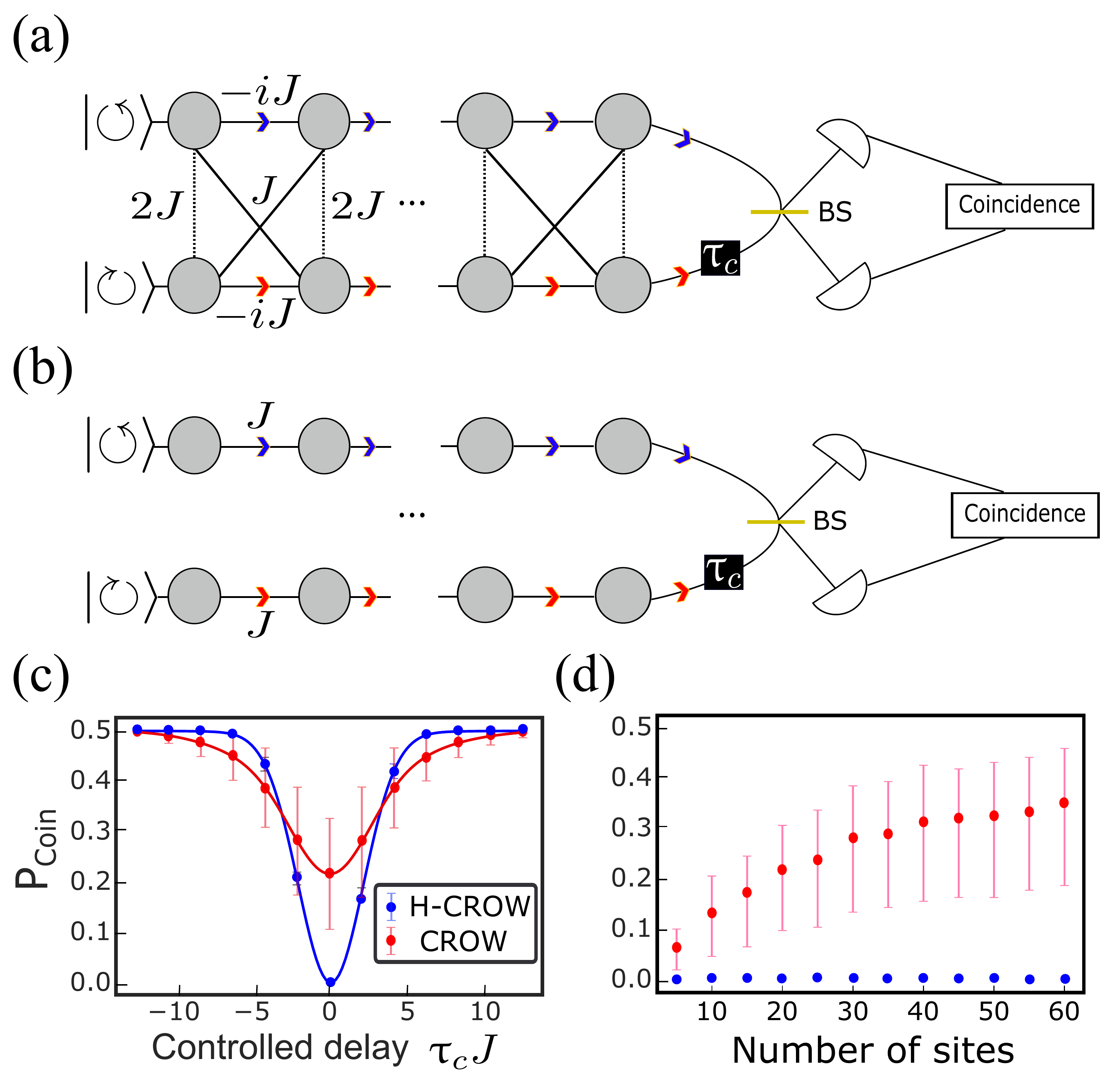}
\caption{(a,b) Schematics of coincidence measurement using tight-binding models of an H-CROW and a pair of regular CROWs. We consider measurements for two photons exhibiting opposite helicity with controlled delay time $\tau_{c}$ before a 50:50 beam splitter (BS) and resulting coincidence probability of two photons to produce simultaneous 'clicks' of single-photon detectors.
(c) Coincidence vs. controlled delay time for 20-site long CROW structures and (d) minimum coincidence values with respect to the number of sites. Blue solid line and dots represent the average for H-CROWs and red for CROWs. Error bars indicate 65$\%$ confidence interval for 500 disorder realizations.}
\label{fig:visibility}
\end{figure}

We compare a degree of the temporal overlap of the two photons after each one propagates through a different part of the device, by calculating the coincidence probability. It was the first experimental witness of quantum property, as Hong \textit{et al.} showed quantumness by generating entangled photons and measuring their coincidence counts vs. the controlled delay to one of the paths~\cite{Hong88}. When the total time delay is zero, coincidence rates after a 50:50 beam splitter reach a minimum and vanish due to the quantum interference when photons are temporally indistinguishable. 
Accordingly, we analyze the photon interference at the output with a tunable temporal delay, as illustrated in Figs.~\ref{fig:visibility}(a,b).
When a two-photon output state passes a beam splitter with ratio $r:t$, where $r$ and $t$ represent reflection and transmission, respectively, the field operators obey the unitary relation~\cite{Loudon00},
\begin{equation}
    \left( \begin{array}{ccc}
		\hat{c}^{\dagger} \\
		\hat{d}^{\dagger}
		\end{array} \right) = \left( \begin{array}{ccc}
		t & ir \\
		ir & t
		\end{array} \right)\left( \begin{array}{ccc}
		\hat{a}^{\dagger}_{\mathrm{out}} \\
		\hat{b}^{\dagger}_{\mathrm{out}}
		\end{array} \right),
\label{eq:beam_matrix}
\end{equation}
where $t^{2} + r^{2} = 1$ ($t,r \in \mathcal{R}$) and $\hat{c},\hat{d}$ indicate the field operators of upper/lower sides after passing the beam splitter. 
We calculate the coincidence probability of the simultaneous `clicks' with the two single-photon detectors, $P_{\text{coin}}$, using the projection operator $\hat{P}_{c} \otimes \hat{P}_{d} \coloneqq \int d\omega  \hat{c}^{\dagger}(\omega)\vert 0\rangle\langle 0 \vert \hat{c}(\omega)   \otimes \int d\omega' \hat{d}^{\dagger}(\omega') \vert 0\rangle\langle 0 \vert \hat{d}(\omega')$~\cite{Branczyk17}. Thus, $P_{\text{coin}}=\text{Tr}[\rho \hat{P}_{c} \otimes \hat{P}_{d}]$, where the density matrix is $\rho = \vert \text{out}\rangle \langle \text{out} \vert$. 
Now, we introduce a tunable delay parameter between two outputs $\tau_{c}$ which controls the temporal overlap between the photons before their interference on the beam splitter. Without loss of generality, we apply this delay to the lower output port, and obtain a final expression for a balanced beam splitter ($t = r = 1/\sqrt{2}$)~\cite{Titchener18},
\begin{equation}
    P_{\text{coin}}(\tau_{c}) = \frac{1}{2}\left(1 - \frac{\left\vert \int d\omega p^{\ast}_{1}(\omega)p_{2}(\omega)e^{i\omega \tau_{c}}\right\vert^{2}}{\int d\omega \left\vert p_{1}(\omega)\right\vert^{2} \int d\omega' \left\vert p_{2}(\omega')\right\vert^{2}}\right).
\label{eq:coincidence}
\end{equation}
The second part of the above equation defines the visibility $V$, $V = \sqrt{1-2P_{\text{coin}}}$~\cite{Loudon00}, which quantifies the degree of interference. In quantum mechanics, it is also referred to as indistinguishability of photons~\cite{Zych15}. The meaning of coincidence probability is thus the resultant distribution of indistinguishability~\cite{Legero03}. Note that coincidence probability is zero for an ideal case of identical single photons, indicating indistinguishability preservation, since the transformation for indistinguishable photon input state reads~\cite{Zou91,Weihs01}
\begin{equation}
    \vert 11\rangle_{ab} \rightarrow \frac{1}{\sqrt{2}}( \vert 20 \rangle_{ab} + \vert 02 \rangle_{ab} ).
\label{eq:BS 11}
\end{equation}
Conversely, for a very large time delay exceeding the wavepacket temporal width, the photons do not interfere with each other, and the coincidence probability becomes 1/2 for a balanced ($50:50$) beam splitter, corresponding to distinguishable photons according to the calculations in Appendix~\ref{app:coincidence}.

We perform a comparison against regular CROWs with the same input and the same propagation length, as illustrated in Figs.~\ref{fig:visibility}(a,b). We consider wavepackets with a Gaussian envelope at the input, i.e. $p_{\text{in}}(\omega) = \exp(-\omega^{2}/2\sigma^{2})$, where $\sigma = 0.5J$ is the envelope width. We plot in Fig.~\ref{fig:visibility}(c) the calculated coincidence probability vs. controlled time delay after propagation through the 20-ring CROW discussed above. 
We note that the separable two-photon state is insensitive to the relative phase delay accumulated along the different paths as sketched in Fig.~\ref{fig:indistinguishability}(a), but can be affected by disorder-induced variations to the group delay and wavepacket distortions as indicated in Figs.~\ref{fig:indistinguishability}(b,c). We observe that the minimum coincidence probability for H-CROWs remains close to zero ($\approx 2\times10^{-4}$), indicating photon indistinguishability is well-preserved. On the contrary, for regular CROWs, the mean minimum coincidence probability is $0.22$, which means that photons become partially distinguishable due to disorder. 
Even though the two photons with opposite spins travel along different disordered paths, the temporal shape of photons at the H-CROW outputs remains almost identical or indistinguishable, as a result of disorder-resistant transport. This advantage becomes more pronounced for longer CROWs, as shown in Fig.~\ref{fig:visibility}(d). The minimum coincidence for H-CROWs remains small ($\approx 4\times10^{-3}$), whereas the minimum coincidence probability for the regular CROWs keeps increasing, indicating reduced indistinguishability as the length of the delay line increases.

\section{Protection of photon entanglement} \label{sec:entang}

We now aim to show that H-CROWs can preserve a peculiar quantum property of transmitted photons, entanglement inherently originating from the quantum coherence. Let us consider a N00N state as an input, $|N::N\rangle \coloneqq (\vert N0 \rangle_{ab} + \vert 0N \rangle_{ab} )/ \sqrt{2}$. Such states are strongly sensitive to all effects of the disorder as sketched in Fig.~\ref{fig:indistinguishability} including phase fluctuations~\cite{Lee02}, in contrast to the separable states we analyzed in the previous section.
The corresponding output state in the frequency domain $\vert \text{out} \rangle$ is
\begin{equation}
\begin{aligned}
|\text{out}\rangle  &= \frac{1}{\sqrt{2}\sqrt{N!}} \left(\hat{\phi}^{N}_{a} + \hat{\phi}^{N}_{b} \right)\vert 00\rangle_{ab} \\& = \frac{1}{\sqrt{2N!}}\left(\int \prod_{i=1}^{N} d\omega_{i}  p_{1}(\omega_{i})\hat{a}_{\mathrm{out}}^{\dagger}(\omega_{i}) \right. \\& \left. \hspace{5em} + \int \prod_{i=1}^{N} d\omega_{i} p_{2}(\omega_{i})\hat{b}_{\mathrm{out}}^{\dagger}(\omega_{i})\right) \vert 00\rangle_{ab} ,
\label{eq:2002 state}
\end{aligned}
\end{equation}
where $\hat{\phi}_{j}$ are the output field operators introduced in the previous section. Let us consider the two-photon state with $N = 2$, which can be simply created from a separable state by passing it through a balanced beam splitter before coupling into the CROW. The output state, after applying the time delay ($\tau_{c}$) but before the very last interference stage, can be expressed as
\begin{equation}
\begin{aligned}
   |\text{out}(\tau_{c})\rangle_{\text{bef}}  
   &= \frac{1}{2}\int d\omega_{1}d\omega_{2}  \left( p_{1}(\omega_{1})p_{1}(\omega_{2})\hat{a}_{\mathrm{out}}^{\dagger}(\omega_{1})\hat{a}_{\mathrm{out}}^{\dagger}(\omega_{2}) \right. \\& \left. + p_{2}(\omega_{1})p_{2}(\omega_{2})\hat{b}_{\mathrm{out}}^{\dagger}(\omega_{1})\hat{b}_{\mathrm{out}}^{\dagger}(\omega_{2})e^{i(\omega_{1}+\omega_{2})\tau_{c}} \right) \vert 00\rangle_{ab}. 
\end{aligned}
\end{equation}
We first analyze the case of indistinguishable photons at the output, and note that in this ideal situations the disorder can introduce a phase difference $\theta$ between the photon in two output ports. The transformation of such a state by output 50:50 beam splitter implements a reversed Hong-Ou-Mandel (HOM) interference~\cite{Paesani20}, which we express as follows:
\begin{equation}
\begin{aligned}
     &\frac{1}{\sqrt{2}}( \vert 20 \rangle_{ab} + e^{i \theta}\vert 02 \rangle_{ab} )\rightarrow  \\& \frac{1}{2\sqrt{2}}\left( (1-e^{i\theta})\vert 20 \rangle_{ab} +\sqrt{2}i(1+e^{i\theta})\vert 11 \rangle_{ab}- (1-e^{i \theta})\vert 02 \rangle_{ab} \right).
\end{aligned}
\label{eq:BS 2002}
\end{equation}
The coincidence probability is then
\begin{equation}
    P_{\text{coin}} = \frac{1+\cos \theta}{2}.
\label{eq:coincidence_fock}
\end{equation}
Note that the coincidence probability can oscillate even though the entangled state remains pure. It is due to the phase sensitivity of the N00N state~\cite{Slussarenko17}. 

Next, we determine the coincidence probability in the general case, taking into account all the effects due to disorder. We compute the density matrix of the output state $\rho = \vert \text{out} \rangle \langle \text{out} \vert$ using the projection operators, 
\begin{equation}
\begin{aligned}
&\underaccent{ab}{|20\rangle\langle20|} = \frac{1}{2}\int d\omega_{1} d\omega_{2}   \hat{a}_{\mathrm{out}}^{\dagger}(\omega_{1})\hat{a}_{\mathrm{out}}^{\dagger}(\omega_{2})\underaccent{ab}{|00\rangle\langle00|}\hat{a}_{\mathrm{out}}(\omega_{2}) \hat{a}_{\mathrm{out}}(\omega_{1}), 
\\& \underaccent{ab}{|02\rangle\langle02|} = \frac{1}{2} \int d\omega_1 d\omega_2 \hat{b}_{\mathrm{out}}^{\dagger}(\omega_{1})\hat{b}_{\mathrm{out}}^{\dagger}(\omega_{2})\underaccent{ab}{|00\rangle\langle00|}\hat{b}_{\mathrm{out}}(\omega_{2}) \hat{b}_{\mathrm{out}}(\omega_{1}), \\& \underaccent{ab}{|11\rangle\langle11|} =  \int d\omega_1 d\omega_2 \hat{a}_{\mathrm{out}}^{\dagger}(\omega_{1})\hat{b}_{\mathrm{out}}^{\dagger}(\omega_{2})\underaccent{ab}{|00\rangle\langle00|}\hat{b}_{\mathrm{out}}(\omega_{2}) \hat{a}_{\mathrm{out}}(\omega_{1}),
\end{aligned}
\end{equation}
respectively. Again, $\hat{a}_{\mathrm{out}}, \hat{b}_{\mathrm{out}}$ denote annihilation operators on upper and lower output legs, respectively. The coincidence probability with the normalized output $\text{Tr}[\rho] = 1$ is given by
\begin{equation}
\begin{aligned}
    &P_{\text{coin}}(\tau_{c}) =\frac{1}{2}  \left(1 + \frac{\left\lbrace\int d\omega  p^{\ast}_{1}(\omega)p_{2}(\omega) e^{i\omega \tau_{c}} \right\rbrace^{2}+\text{c.c.}}{\left\lbrace\int d\omega  \left\vert p_{1}(\omega)\right\vert^{2} \right\rbrace^{2}+\left\lbrace\int d\omega  \left\vert p_{2}(\omega)\right\vert^{2} \right\rbrace^{2}}\right).
\label{eq:entanglement coincidence}
\end{aligned}
\end{equation}
In agreement with the expression written in Fock basis in~\eqref{eq:coincidence_fock}, here phase fluctuations $\theta$ arise from phase mismatches between the fields $p_{1,2}(\omega)$. Note that the coincidence probability is ${}_{ab}\langle 11\vert \rho \vert 11 \rangle_{ab}$, which we derive in Appendix~\ref{app:density2002}. 

To quantify the mixedness of the output state induced by disorder, we also analyze another quantity, the purity $\text{Tr}[\rho^{2}]$, which is bounded by $1/d \leq \text{Tr}[\rho^{2}] \leq 1$, where $d$ is the dimension of Hilbert space, i.e. $d = 2$ for the two-photon case. The maximum value corresponds to pure states, and the minimum to fully mixed states. The state purity after passing the controlled delay is
\begin{equation}
\begin{aligned}
    &\text{Tr}[\rho^{2}(\tau_{c})] = 1 \\
    &+ 2\left(\frac{  \left\lbrace\left\vert\int d\omega  p^{\ast}_{\text{1}}p_{\text{2}} e^{i\omega \tau_{c}}\right\vert^{2} \right\rbrace^{2}  - \left\lbrace\int d\omega  \left\vert p_{\text{1}}\right\vert^{2} \int d\omega'\left\vert p_{\text{2}}\right\vert^{2} \right\rbrace^{2}}{\left(\left\lbrace\int d\omega  \left\vert p_{\text{1}}\right\vert^{2} \right\rbrace^{2}+\left\lbrace\int d\omega  \left\vert p_{\text{2}}\right\vert^{2} \right\rbrace^{2}\right)^{2}}\right),
\label{eq:entanglement purity}
\end{aligned}
\end{equation}
where we omit the integral variable $\omega,\omega'$ to simplify the notation.

The form of Eq.~(\ref{eq:entanglement purity}) reveals that the output state remains pure when indistinguishability is preserved. Namely, since we consider non-interacting particles we can relate the N00N state purity to the coincidence probability of the separable input state $\vert 11\rangle_{ab}$ according to \eqref{eq:coincidence}. Since $P_{\mathrm{coin}}(\tau_c) \approx 0$, we have  $\left\vert\left\lbrace\int d\omega  p^{\ast}_{\text{1}}p_{\text{2}} e^{i\omega \tau_{c}} \right\rbrace \right\vert^{2} \approx \left\lbrace\int d\omega  \left\vert p_{\text{1}}\right\vert^{2} \right\rbrace \left\lbrace\int d\omega'\left\vert p_{\text{2}}\right\vert^{2} \right\rbrace$, indicating that the two output fields $p_{1}, p_{2}$ have identical intensities and group delays. The expression in \eqref{eq:entanglement purity} thereby approaches 1 even though coincidence probability for $\vert 2::2 \rangle$ may fluctuate due to the phase sensitivity of N00N state.

\begin{figure}
\centering
\includegraphics[width=\columnwidth,height=6.0cm]{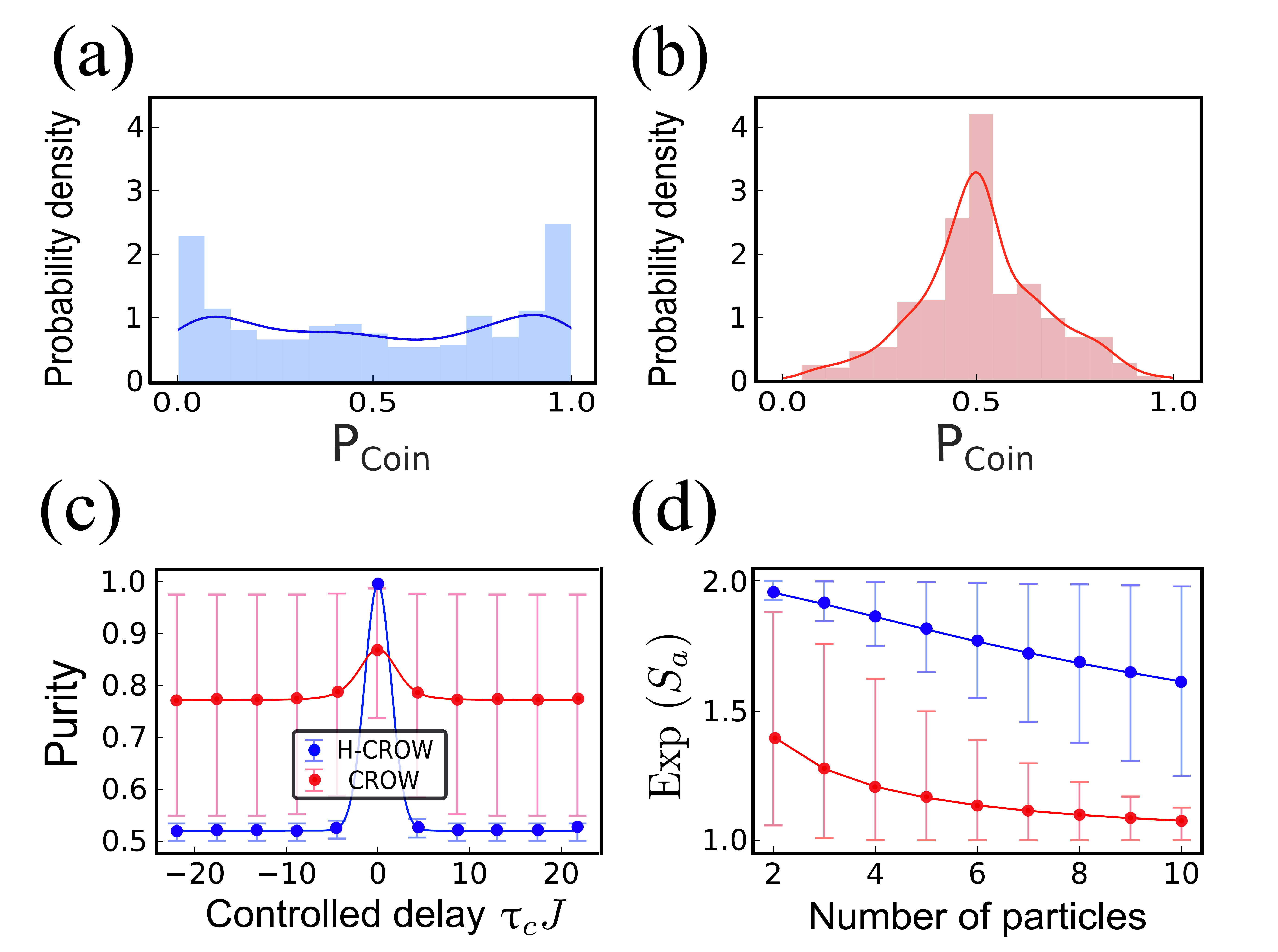}
\caption{Disorder-robust transmission of $N=2$ N00N states using the H-CROW. (a,b) Statistics of the output coincidence probability for the H-CROW (a), and regular CROWs (b). (c) Output state purity for the H-CROW (blue) and regular CROW (red), with error bars indicating 65\% confidence interval. (d) Exponentiated entanglement entropy of the upper output port, $\exp(S_{a})$, which distinguishes maximally entangled states $\exp(S_a) = 2$ from separable states $\exp(S_a) = 1$. We use an ensemble of 500 disorder realizations and disorder strength $U= 0.8J$.}
\label{fig:entangled}
\end{figure}

To verify this reasoning, we plot in Figs.~\ref{fig:entangled}(a,b) the distributions of coincidence probability for N00N states transmitted through H-CROWs and regular CROWs, respectively, with a zero controlled time delay ($\tau_{c} = 0$). We observe that for H-CROWs the oscillation of probability occurs in the full range of $[0,1]$, which is evidence that entanglement is preserved. In contrast, the coincidences from the regular CROWs show a peak at 0.5, indicating the output state is mixed and the entanglement is lost due to disorder. We additionally show in Fig.~\ref{fig:entangled}(c) the average purity and the 65\% confidence interval vs. the delay time. We see that purity stays at one for H-CROWs at zero delay, while CROWs exhibit loss coherence with huge fluctuations due to disorder (red). 

To further quantify the effect of disorder, we consider the entanglement entropy $S_a$, which indicates the capacity for encoding quantum information~\cite{Nielsen02},
\begin{equation}
S_{a} \coloneqq -\text{Tr}_{a}[\rho_{a} \text{ln}(\rho_{a})],
\label{eq:entanglement entropy}    
\end{equation}
where $\rho_{a} = \text{Tr}_{b}[\rho(\tau_{c})]$. Note that as coherence terms vanish by partial trace, the controlled delay does not affect the entropy. We compute the entanglement entropy in Appendix~\ref{app:density2002},
\begin{equation}
\begin{aligned}
    S_{a} = -x\text{ln}x - (1-x)\text{ln}(1-x),
\end{aligned}
\label{eq:entanglement entropy expression}
\end{equation}
where for the two-photon ($N = 2$) N00N state we have
\begin{equation}
   x = \frac{\left\lbrace\int d\omega  \left\vert p_{\text{out,1}}(\omega)\right\vert^{2} \right\rbrace^{2}}{\left\lbrace\int d\omega  \left\vert p_{\text{out,1}}(\omega)\right\vert^{2} \right\rbrace^{2} + \left\lbrace\int d\omega  \left\vert p_{\text{out,2}}(\omega)\right\vert^{2} \right\rbrace^{2} }.
\end{equation}
The entanglement is maximized when $x = 0.5$, corresponding to identical intensities at the two output ports. This demonstrates the higher capacity of H-CROWs, which exhibit almost identical output intensities from the upper and lower ports. In contrast, CROWs yield poor capacity as the intensity ratio differs due to disorder. We show in Fig.~\ref{fig:entangled}(d) the scaling of the entanglement entropy with the number of photons, revealing huge fluctuations and loss of entanglement for the regular CROW, while the H-CROW can preserve some amount of entanglement.

\section{Conclusion} \label{sec:concl}

In this manuscript, we have studied the propagation of quantum states of light through helical coupled-resonator waveguides (H-CROWs). Regular CROWs can serve as delay lines in integrated photonic circuits, however, they exhibit a strong sensitivity to fabrication disorder preventing reliable transmission of wavepackets. H-CROWs exploit an additional sublattice degree of freedom to achieve disorder-resistant transport, which arises due one-way modes at the center of their transmission band, whose propagation direction is fixed by the excited sublattice (known as pseudospin-momentum locking). Using numerical solutions of tight-binding models describing H-CROWs and regular CROWs, we have shown that the former can be used to more reliably transport quantum states of light in the presence of disorder. 

We firstly showed that transmission probability and wavepacket delay times have narrower fluctuations and provide more ballistic-like transport compared to regular CROWs. Next, we showed that two identical photons transmitted through an H-CROW can preserve the indistinguishability of their temporal wavepackets, and accordingly demonstrate the quantum Hong-Ou-Mandel interference. Finally, we showed that path-entangled two-photon N00N states are preserved as pure entangled states, while the effect of disorder is only expressed through the accumulation of a relative phase between the photon pairs. We note that this relative phase fluctuation can be compensated by simply placing a single tunable phase shifter at one of the output ports~\cite{Canciamilla10}. The H-CROWs perform better at reliably transporting both types of quantum states, while quantum features are strongly suppressed by disorder in regular CROWs. 

In the future, it will be interesting to generalize our findings to multi-mode entangled states and multi-mode H-CROWs.
We expect that our results can provide a practical way to create robust integrated photonic delay lines, which can serve as essential components facilitating reliable generation and guiding of the quantum state of light for multiple applications, including scalable quantum information processing. 

\section{Acknowledgements}

This research was supported by the Institute for Basic Science in Korea (IBS-R024-Y1). A.A.S. acknowledges support by the Australian Research Council (ARC) (DP190100277).

\begin{appendix}
\section{Input-Output relation} \label{app:in-out}
When we induce the coupling between an external probe waveguide characterized by field operator $\hat{p}$ and the system, the dynamics are governed by total Hamiltonian,
\begin{equation}
    \hat{H}_{\text{tot}} = \hat{H}_{\text{sys}} + \hat{H}_{\text{env}} + \hat{H}_{\text{int}},
\end{equation}
where $\hat{H}_{\text{sys}}$ is our Hamiltonian of interest. Hamiltonians describing the environment $\hat{H}_{\text{env}}$ and the interaction between system and environment $\hat{H}_{\text{int}}$ are given by
\begin{equation}
\begin{aligned}
&\hat{H}_{\text{env}} = \int_{-\infty}^{\infty} d\omega \ \omega \hat{p}^{\dagger}(\omega,t) \hat{p}(\omega,t),
\\&\hat{H}_{\text{int}} = i \int_{-\infty}^{\infty} d\omega  \sqrt{2\kappa_{\text{ex}}} \left(\hat{a}^{\dagger}(t)\hat{p}(\omega,t) - \text{h.c.}\right),
\end{aligned}
\end{equation}
where $\kappa_{\text{ex}}$ is an external coupling parameter independent of frequency, $\hat{a}$ is the coupling field operator of system, e.g. $\hat{a}_{1}$ for upper input port. Note that $\hat{p}_{in}$ lives in a different Hilbert space from the resonator field operators $\hat{a}_{j}$ as it represents the field operator of environment. Besides, negative frequencies are allowed as we work in a rotating frame at a frequency much larger than typical bandwidths we consider~\cite{Gardiner85}. Here we assume that probe waveguide has an almost continuous spectrum. Heisenberg equation of the bath operator reads~\cite{Gardiner85} 
\begin{equation}
    \frac{d}{dt}\hat{p}(\omega,t) = -i\omega  \hat{p}(\omega,t) - \sqrt{2\kappa_{\text{ex}}} \hat{a}(t).
\end{equation}
The solution of the above is
\begin{equation}
    \hat{p}(\omega,t) = e^{-i\omega (t-t_{0})} \hat{p}(\omega,t_{0}) - \sqrt{2\kappa_{\text{ex}}} \int_{t_{0}}^{t} dt' e^{-i\omega (t-t')}\hat{a}(t'),
\label{eq:Langevin 1}
\end{equation}
where $t_{0}$ is the initial time. Here the input field $\hat{p}_{\text{in}}(t)$ is defined as~\cite{Gardiner85}
\begin{equation}
    \hat{p}_{\text{in}}(t) \coloneqq \frac{1}{\sqrt{2\pi}} \int_{-\infty}^{\infty} \ d\omega e^{-i\omega (t-t_{0})} \hat{p}(\omega, t_{0}) \quad (t > t_{0}),
\label{eq:input}
\end{equation}
this is a Fourier transform of the input spectrum. Then,~\eqref{eq:Langevin 1} can be written differently,
\begin{equation}
\begin{aligned}
    \int d\omega \hat{p}(\omega,t) &= \hat{p}_{\text{in}}(t) - \sqrt{2\kappa_{\text{ex}}} \int_{-\infty}^{\infty} d\omega \ \int_{t_{0}}^{t} dt' e^{-i\omega (t-t')}\hat{a}(t') \\& = \hat{p}_{\text{in}}(t) - \frac{\sqrt{2\kappa_{\text{ex}}}}{2} \hat{a}(t),
\end{aligned}
\label{eq:Langevin input}
\end{equation}
here we use $\Theta(t) \coloneqq \int_{-\infty}^{t} dt' \delta(t'), \Theta(0) = 1/2$. We can introduce another solution by defining final time $t_{1}$, it yields the solution, 
\begin{equation}
    \hat{p}(\omega,t) = e^{-i\omega (t-t_{1})} \hat{p}(\omega,t_{1}) - \sqrt{2\kappa_{\text{ex}}} \int_{t_{1}}^{t} dt' e^{-i\omega (t-t')}\hat{a}(t').
\label{eq:Langevin 2}
\end{equation}
Let us define output field operator $\hat{p}_{\text{out}}$ where
\begin{equation}
    \hat{p}_{\text{out}}(t) \coloneqq \frac{1}{\sqrt{2\pi}} \int \ d\omega e^{-i\omega (t-t_{1})} \hat{p}(\omega, t_{1})  \quad (t < t_{1}),
\label{eq:output}
\end{equation}
which yields
\begin{equation}
\begin{aligned}
    \int d\omega \hat{p}(\omega,t) &= \hat{p}_{\text{out}} + \sqrt{2\kappa_{\text{ex}}} \int_{-\infty}^{\infty} d\omega \ \int_{t}^{t_{1}} dt' e^{-i\omega (t-t')}\hat{a}(t') \\& = \hat{p}_{\text{out}}(t) + \frac{\sqrt{2\kappa_{\text{ex}}}}{2} \hat{a}(t).
\end{aligned}
\label{eq:Langevin output}
\end{equation}
We can hence find the identity between input and output~\cite{Gardiner85},
\begin{equation}
    \hat{p}_{\text{in}}(t) - \hat{p}_{\text{out}}(t) = \sqrt{2\kappa_{\text{ex}}}\hat{a}(t),
    \label{in-output identity}  
\end{equation}
which yields the representations of reflection and transmission coefficients described in~\eqref{eq:reflection and transmission}.

\section{Coincidence probability for general beam splitter}  \label{app:coincidence}
Suppose identical photons come toward a beam splitter (BS) from the opposite ports as depicted in Figs.~\ref{fig:visibility}(a,b). BS exhibits the ratio $r:t$, where r,~t~($\in \mathcal{R}$) are the reflectivity and transmissivity, respectively, $r^{2}+t^{2} = 1$. The transformation for indistinguishable photons reads~\cite{Zou91,Weihs01}
\begin{equation}
    \vert 11\rangle_{ab} \rightarrow irt \sqrt{2} \vert 20 \rangle_{ab} +   irt \sqrt{2} \vert 02 \rangle_{ab} + (t^{2}-r^{2}) \vert 11 \rangle_{ab}, 
\label{eq:BS}
\end{equation}
where modes $a$ and $b$ indicate the photons coming from upper and lower parts, respectively. The coincidence probability is now given by $P_{\text{coin}} = (t^{2}-r^{2})^{2}$. Minimum coincidence occurs when $r=t = 1/\sqrt{2}$, yielding $0$ coincidence. On the other hand, given two photons are distinguishable~\eqref{eq:BS} is then
\begin{equation}
\begin{aligned}
    \vert 1001\rangle_{a_{1}b_{1}a_{2}b_{2}} &\rightarrow irt \vert 1010 \rangle_{a_{1}b_{1}a_{2}b_{2}} +   irt \vert 0101 \rangle_{a_{1}b_{1}a_{2}b_{2}} \\&+ t^{2}\vert 1001\rangle_{a_{1}b_{1}a_{2}b_{2}} - r^{2}\vert 0110\rangle_{a_{1}b_{1}a_{2}b_{2}}, 
\label{eq:BS2}
\end{aligned}
\end{equation}
and it turns out that coincidence probability of measuring photons on each detector simultaneously is $P_{\text{coin}} = t^{4} + r^{4}$. Then, $0.5$ is the lowest value possible for a classical field when the splitting ratio is $50:50$. Hence $P_{\text{coin}} = 0.5$ implies the boundary between quantum and classical cases, where a lower value is only possible due to quantum interference between (partially) indistinguishable photons~\cite{Loudon00,Weihs01,Sadana19}.

Similar to our analysis for $\vert11\rangle$ state, we calculate the transformation of the $\vert 2::2\rangle$ state after a beam splitter,
\begin{equation}
\begin{aligned}
    &\frac{1}{\sqrt{2}}\left(\vert 20\rangle_{ab} + \vert 02\rangle_{ab}\right) \rightarrow \\& \frac{1}{\sqrt{2}}\left((t^{2}-r^{2}) \vert 20 \rangle_{ab}- (t^{2}-r^{2}) \vert 02 \rangle_{ab} + 2\sqrt{2}irt \vert 11 \rangle_{ab}\right). 
\end{aligned}
\label{eq:BS3}
\end{equation}
For distinguishable particles, the beam splitter transformation is
\begin{equation}
\begin{aligned}
    &\frac{1}{\sqrt{2}}\left(\vert 1010\rangle_{a_{1}b_{1}a_{2}b_{2}} + \vert 0101\rangle_{a_{1}b_{1}a_{2}b_{2}}\right) \rightarrow \\& \frac{1}{\sqrt{2}}\left((t^{2}-r^{2}) \vert 1010 \rangle_{a_{1}b_{1}a_{2}b_{2}}-    (t^{2}-r^{2}) \vert 0101 \rangle_{a_{1}b_{1}a_{2}b_{2}} \right. \\& \left. + 2irt \vert 0110 \rangle_{a_{1}b_{1}a_{2}b_{2}} + 2irt \vert 1001 \rangle_{a_{1}b_{1}a_{2}b_{2}}\right). 
\end{aligned}
\label{eq:BS4}
\end{equation}
When a balanced beam splitter is used, coincidence probability becomes 1 for both cases, but note that one has which-way information and other does not. Then, coincidence probability does not give enough information on distinguishability, so another measure of entanglement should be taken into account such as entanglement entropy, as we discuss in Sec.~\ref{sec:entang}. 

\section{Density matrix of 2002 state}  \label{app:density2002}

In order to compute coincidence probability and entanglement entropy, one has to introduce a density matrix defined as an outer product of output states. Density matrix of output state for a $\vert 2::2\rangle$ input in terms of Fock basis has the form with normalization,
\begin{equation}
    \rho_{\text{bef}}(\tau_{c}) = \frac{1}{A+B}\left( \begin{array}{cc}
		A & C(\tau_{c})  \\
		C^{\ast}(\tau_{c}) & B 	\end{array} \right),
\end{equation}
where
\begin{equation}
\begin{aligned}
&A = {}_{ab}\langle 20 \vert \rho_{\text{bef}}(\tau_{c}) \vert 20 \rangle_{ab} \coloneqq \rho_{\text{bef,2020}}= 
		\left\lbrace\int d\omega  \left\vert p_{\text{out,1}}(\omega)\right\vert^{2} \right\rbrace^{2}, \\&B  = \rho_{\text{bef,0202}} = \left\lbrace\int d\omega  \left\vert p_{\text{out,2}}(\omega)\right\vert^{2} \right\rbrace^{2}, \\& C(\tau_{c})  = \rho_{\text{bef,2002}} = \left\lbrace\int d\omega  p^{\ast}_{\text{out,1}}(\omega)p_{\text{out,2}}(\omega) e^{i\omega \tau_{c}} \right\rbrace^{2} .
\end{aligned}
\end{equation}
Note that coincidence probability ($\langle 11 \vert \rho_{\text{bef}} \vert 11\rangle$) is zero before it passes through the last beam splitter since the modes are decoupled in our system. After the photon state exhibits interference in the last beam splitter, the density matrix $\rho_{\text{af}}(\tau_{c})$ becomes
\begin{equation}
\begin{aligned}
&\rho_{\text{af},2020} = \rho_{\text{af},0202} = -\rho_{\text{af},2002} = -\rho_{\text{af},0220} \\&= \frac{1}{4(A+B)} \left[A+B-C(\tau_{c})-C^{\ast}(\tau_{c})\right],\\& \rho_{\text{af},2011} = \rho_{\text{af},1120}^{\ast} = -\rho_{\text{af},0211} = -\rho_{\text{af},1102}^{\ast} \\&= \frac{i}{2\sqrt{2}(A+B)}\left[A-B+C(\tau_{c})-C^{\ast}(\tau_{c})\right], \\& \rho_{\text{af},1111} =\frac{1}{2(A+B)}  \left[A+B+C(\tau_{c})+C^{\ast}(\tau_{c})\right],
\label{eq:density_after}
\end{aligned}
\end{equation}
where each component labels the basis $\lbrace \vert 20\rangle_{ab}, \vert 11\rangle_{ab}, \vert 02\rangle_{ab} \rbrace$. Note that the last term of above equations indicates the coincidence probability~\eqref{eq:entanglement coincidence}. This expression can be obtained equivalently using the unitary operator in operator basis we have defined in~\eqref{eq:beam_matrix}. The unitary matrix for two particles in the tensor product form reads
\begin{equation}
    U\otimes U \coloneqq U_{2} = \frac{1}{2}\left( \begin{array}{cccc}
		1 & i & i & -1 \\
		i & 1 & -1 & i \\
		i & -1 & 1 & i \\
		-1 & i & i & 1
		\end{array} \right),
\end{equation}
as we consider indistinguishable particles, it can be contracted on the basis of $\lbrace |20\rangle_{ab} , |11\rangle_{ab}, |02\rangle_{ab} \rbrace$,
\begin{equation}
    U_{2} = \frac{1}{2}\left( \begin{array}{ccc}
		1 & \sqrt{2}i & -1 \\
		\sqrt{2}i & 0 & \sqrt{2}i \\
		-1 & \sqrt{2}i & 1 
		\end{array} \right).
\end{equation}
One can check that~\eqref{eq:density_after} is equivalent to $\rho_{\text{af}}(\tau_{c}) = U_{2}\rho_{\text{bef}}(\tau_{c})U^{\dagger}_{2}$. The purity  $\text{Tr}[\rho^{2}]$ thus has the form
\begin{equation}
\begin{aligned}
    \text{Tr}[\rho^{2}(\tau_{c})] &= \frac{\left( A ^{2} + B ^{2} + 2\vert C(\tau_{c}) \vert^{2}  \right)}{(A+B)^2} = 1 +2 \frac{\vert C(\tau_{c}) \vert^{2}-AB}{(A+B)^2}.
\label{eq:purity_bef_norm}
\end{aligned}
\end{equation}
Note that the purity expressions are identical for $\rho_{\text{bef}}$ and $\rho_{\text{af}}$ due to the property of unitary transform. Positivity of each term in numerator guarantees positivity of purity, and it turns out to be equal or less than unity because of H\"{o}lder's inequality, $AB \geq \vert C(\tau_{c}) \vert^{2}$~\cite{Faria10}.

Entanglement entropy can be obtained from the above ingredients. Tracing out the lower port degree of freedom ($b$) yields the expression about the reduced density matrix $\rho_{a}$,
\begin{equation}
\begin{aligned}
    S_{a} &= -\text{Tr}_{a}[\rho_{a}\text{ln}(\rho_{a})]
    = -\frac{A}{A+B}\text{ln}\frac{A}{A+B} - \frac{B}{A+B}\text{ln}\frac{B}{A+B}.
\end{aligned}
\label{eq:entanglement entropy expression appendix}
\end{equation}
Note that the controlled delay does not affect the entanglement entropy as off-diagonal components $C,C^{\ast}$ representing phase mismatches between fields do not play any role, while the only relative intensity of two ports determines this entanglement entropy.
\end{appendix}

\bigskip
\noindent 

\bibliography{ref}

\bibliographyfullrefs{ref}

\end{document}